\documentclass[sigconf]{acmart}

\copyrightyear{2026}
\acmYear{2026}
\setcopyright{cc}
\setcctype{by}

\acmISBN{979-8-4007-2539-5/2026/11}
\acmDOI{10.1145/3799682.3840961}

\acmConference[CIKM '26]{The 35th ACM International Conference on Information and Knowledge Management}{November 7--11, 2026}{Rome, Italy}
\acmBooktitle{Proceedings of the 35th ACM International Conference on Information and Knowledge Management (CIKM '26)}

\makeatletter
\patchcmd{\maketitle}
  {\itshape \acmConference@shortname, \acmConference@venue}
  {\itshape \acmConference@shortname, \acmConference@date, \acmConference@venue}
  {}{\PackageError{bridge}{Could not add the CIKM conference date to the copyright block}{Check the acmart version and maketitle patch.}}
\makeatother

\AtBeginDocument{%
  }

\begin{document}

\title{BRIDGE: Behavior-Guided Residual Integration with Dual-Frequency Graph Evidence}

\author{Zesheng Li}
\affiliation{%
  \institution{University of the Chinese Academy of Sciences}
  \city{Beijing}
  \country{China}}
\email{2022111515@stu.sufe.edu.cn}

\author{Chengchang Pan}
\authornote{Corresponding authors.}
\affiliation{%
  \institution{University of the Chinese Academy of Sciences}
  \city{Beijing}
  \country{China}}
\email{166353314@qq.com}

\author{Honggang Qi}
\authornotemark[1]
\affiliation{%
  \institution{University of the Chinese Academy of Sciences}
  \city{Beijing}
  \country{China}}
\email{hgqi@ucas.ac.cn}

\renewcommand{\shortauthors}{Li et al.}

\begin{abstract}
Multimodal recommendation improves item representations by combining
visual, textual, and collaborative signals, but stronger cross-view
alignment does not always improve ranking. Our diagnostics on Amazon
Baby show that direct consistency has an effective range: moderate
alignment helps, while stronger alignment suppresses recommendation-specific
variation. We also observe a clear spectral split: low-frequency
components capture shared structure, whereas higher-frequency
components retain more private ranking signal. Based on these findings,
we propose BRIDGE, a behavior-guided residual integration framework
built on dual-frequency graph evidence. BRIDGE separates the model into
three parts: DFGE decomposes graph-smoothed ID, visual, and textual
channels into spectral bands; BEN converts training-only co-user
overlap into signed candidate evidence; and CRI applies that evidence
only inside the base top-$K$ candidate set during training and
inference. This design keeps the multimodal backbone and localizes
behavior evidence to candidate calibration. Experiments on
Amazon Baby, Sports, and Electronics show that BRIDGE reaches
0.1128/0.1262/0.0778 Recall@20 and 0.0525/0.0594/0.0385 NDCG@20,
outperforming baselines by up to 7.3\% in Recall@20 and 14.9\% in
NDCG@20. Project materials are available at
\url{https://lizesheng13.github.io/bridge/}.
\end{abstract}



\maketitle

\section{Introduction}

Images, titles, descriptions, and brands provide useful side
information for recommendation, especially when interaction data are
sparse. This has motivated a broad family of multimodal recommenders
that combine user-item interactions with visual and textual item
features \cite{he2016vbpr,wei2019mmgcn,wei2020grcn,wang2021dualgnn}.
Recent surveys synthesize this literature from representation, modeling,
and optimization perspectives \cite{pan2026mrsurvey}. Recent work has
improved the representation layer by building item-item graphs from
content, denoising multimodal features, or enforcing stronger
consistency across views \cite{tao2022slmrec,zhou2023bm3,
jiang2024diffmm,liu2024alignrec,yang2025fitmm}. The common premise is
simple: a better multimodal representation should yield a better ranker.

Our diagnostics point to a narrower role for content evidence. Moderate
cross-view consistency improves ranking, whereas stronger consistency
suppresses recommendation-specific variation. Frequency analysis shows a
shared-vs-private split: low-frequency bands agree across views, while
higher-frequency bands retain more ranking signal. This distinction is
most important inside the small candidate set where ranking is decided.
Over-aligning the views blurs the differences between plausible items
and near neighbors.

BRIDGE turns these observations into a behavior-guided design. The
multimodal encoder defines the base candidate geometry, and co-user
behavior authorizes only local residual corrections inside that
candidate set. DFGE preserves frequency-aware multimodal evidence, BEN
converts training-only co-user overlap into signed behavior support,
and CRI applies that support only within the candidate scope. On Amazon
Baby, Sports, and Electronics, BRIDGE reaches 0.1128, 0.1262, and
0.0778 Recall@20, respectively, with corresponding NDCG@20 values of
0.0525, 0.0594, and 0.0385.

The deployment setting adds a second constraint. In a retrieval-and-rank
pipeline, the shortlist is the decision bottleneck. A content signal
that helps representation learning can hurt ranking when it moves every
catalog score. BRIDGE keeps the multimodal backbone responsible for the
shortlist and lets behavior evidence revise only candidates that the
backbone has already made plausible.

The contributions of this work are:
\begin{itemize}
  \item We identify two diagnostics in multimodal recommendation:
  direct cross-view consistency has an effective range, and the
  representation spectrum separates shared low-frequency evidence from
  more private higher-frequency variation.
  \item We propose DFGE, a dual-frequency graph evidence encoder that
  preserves that spectrum with band routing and structural
  regularization.
  \item We propose BEN and CRI, which turn training-only co-user
  overlap into signed behavior evidence and use it only to calibrate
  the candidate scope used at training and inference.
\end{itemize}

\begin{figure*}[!t]
  \centering
  \includegraphics[width=0.92\textwidth]{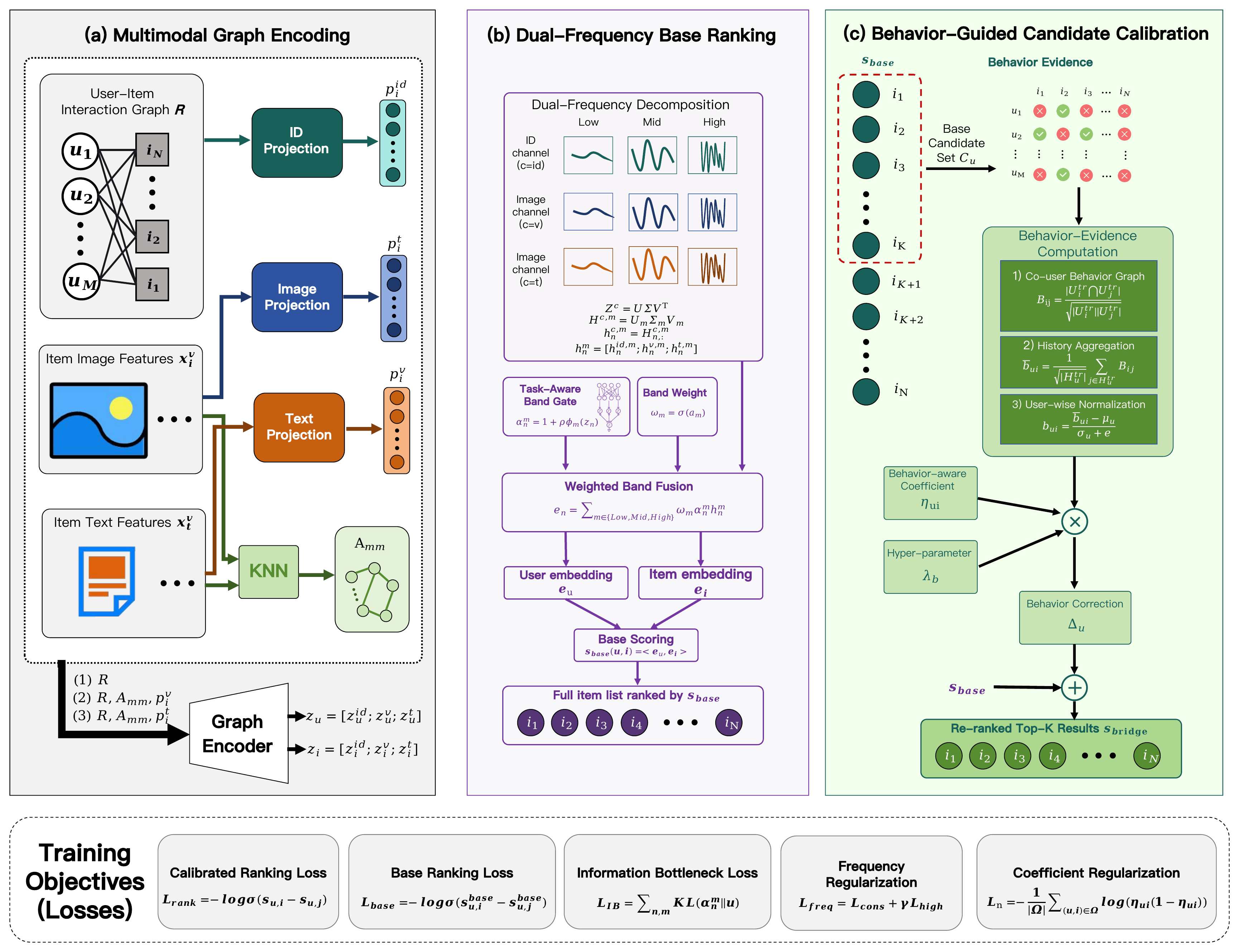}
  \caption{Overview of BRIDGE. DFGE constructs dual-frequency graph
  evidence; BEN computes normalized behavior support; CRI integrates
  the residual only inside the selected candidate set.}
  \Description{A three-stage BRIDGE pipeline. The left panel encodes the
  user-item graph together with image and text item features and builds a
  content KNN graph. The center panel decomposes ID, image, and text
  channels into low, middle, and high frequency bands, then fuses them
  into base user and item representations and ranks the full item list.
  The right panel selects the base top-K candidates, computes co-user
  behavior evidence, and applies a weighted residual only to those
  candidates before producing the final ranking. A bottom strip lists
  the calibrated ranking, base ranking, information bottleneck,
  frequency regularization, and coefficient regularization losses.}
  \label{fig:overview}
\end{figure*}

\section{Related Work}

\subsection{Multimodal Recommendation and Cross-View Alignment}
Multimodal recommendation extends collaborative filtering with visual
and textual signals. VBPR \cite{he2016vbpr} shows that content can
alleviate sparsity, while graph-based methods such as MMGCN
\cite{wei2019mmgcn}, GRCN \cite{wei2020grcn}, DualGNN
\cite{wang2021dualgnn}, and LATTICE \cite{zhang2021lattice} organize
these signals with user-item and item-item structure. Recent variants
add denoising, diffusion, robustness, dual-path routing, retrieval
completion, or stronger cross-view consistency \cite{tao2022slmrec,
zhou2023freedom,zhou2023bm3,jiang2024diffmm,liu2024alignrec,
ping2024ddrec,zhong2024mirrorgradient,qi2025even,dai2026crane,
guo2026igdmrec,li2026gremc,liu2025dsgrec,lu2026recgoat,li2024whotoalign,
ma2024mcdrec,xiu2026dcarmd}. Related alignment and fusion work also
explores dual-graph attention, hierarchical multi-step fusion, and
multimodal large language model coupling \cite{lei2023tmfun,
yang2024mmil,hua2024mllmmsr,qin2026mifusr,sahyouni2026musicrec}. The
common theme is to improve representation quality before ranking.
BRIDGE differs by asking when content evidence should be trusted by the
ranker rather than how aggressively it should be fused.

Most alignment methods use cross-view agreement as a proxy for better
ranking. When content matching dominates, it can erase the differences
between a plausible item and a near duplicate or popular lookalike.
BRIDGE uses the same multimodal evidence to define candidate geometry,
then decides whether behavior evidence should change the rank of each
candidate. This decision is part of the model rather than a post-hoc
reranking rule.

\subsection{Frequency-Aware and Behavior-Guided Recommendation}
Frequency-aware recommenders separate stable structure from fine-grained
variation. SMORE \cite{ong2024smore}, FITMM \cite{yang2025fitmm}, SSR
\cite{yang2025ssr}, MSCF-Net \cite{li2026mscfnet}, DuGRec
\cite{ma2026dugrec}, D-DPDG \cite{wu2026ddpdg}, and
Sequences-as-Nodes \cite{sahyouni2026musicrec} show that spectral views
can organize multimodal or collaborative evidence. These works differ
in whether the spectrum is used for denoising, co-frequency enhancement,
or explicit band routing. BRIDGE uses the same general intuition but
keeps the band decomposition as a representation core rather than a
standalone predictor.

That makes BRIDGE different from frequency-aware denoising pipelines.
Those models usually use the spectrum to suppress noise or separate
stable from unstable components, while BRIDGE uses the decomposition to
decide which bands remain shared evidence and which bands are left as
residual variation for candidate calibration. The band split is
therefore an authorization step, not only a regularizer.

The same point applies to band routing. A decomposition can denoise a
representation, but denoising alone does not explain how the output
should affect ranking. BRIDGE keeps the spectrum inside the backbone
and lets candidate calibration consume only the bands that remain
useful after decomposition. This keeps the representation and the
calibration rule aligned with one another.

\subsection{Behavior-Guided and Candidate-Side Control}
Item-neighborhood methods rank with interaction-derived item scores,
while linear item--item models learn a global coefficient matrix from
the interaction data \cite{sarwar2001itembased,steck2019ease}. These
approaches are close to the behavior branch in BRIDGE, but they are
normally used as standalone catalog-wide rankers. BRIDGE instead treats
raw co-occurrence and EASE as candidate-side residual controls and
applies signed behavior evidence only after the multimodal backbone has
selected a top-$K$ candidate set.

In parallel, behavior-guided methods such as BPR
\cite{rendle2009bpr}, LightGCN \cite{he2020lightgcn}, DRAGON
\cite{zhou2023dragon}, FGCM \cite{xiao2025fgcm}, JBM-Diff
\cite{pan2026jbmdiff}, and causal-inspired multimodal recommendation
\cite{yang2025causalmmrec} use interaction structure to validate or
filter content-derived relations. More recent work extends this idea to
dual-graph diffusion, modality coherence, and multi-intention or
sequential settings \cite{guo2026igdmrec,ma2024mcdrec,hua2024mllmmsr,
yang2024mmil,qin2026mifusr}. BRIDGE combines these lines,
but uses behavior only as a local authorization signal for candidate-level
residual calibration rather than as a global fusion or denoising
target.

Unlike a post-hoc re-ranker or a catalog-wide item--item score,
BRIDGE trains and evaluates its residual under the same detached top-$K$
scope. The base scorer remains responsible for retrieving new positives,
while behavior evidence changes only the local ordering of plausible
candidates.

Together, these strands motivate BRIDGE's dual-frequency and
candidate-calibrated design.

\section{Problem Formulation}

Let $\mathcal{U}$ and $\mathcal{I}$ denote the user and item sets. We
observe implicit feedback
$\mathcal{R}=\{(u,i)\mid u\in\mathcal{U}, i\in\mathcal{I}\}$, where an
edge indicates that user $u$ interacted with item $i$. Each item also
has multimodal content features, including visual features
$\mathbf{x}^{v}_{i}$ and textual features $\mathbf{x}^{t}_{i}$. For a
user $u$, let $\mathcal{H}_{u}$ be the interacted item set; for an item
$i$, let $\mathcal{U}_{i}$ be the set of users who interacted with it.
The recommendation task is to learn a score $s(u,i)$ and rank
unobserved items for each user under full-sort evaluation.

The standard training objective is pairwise Bayesian Personalized
Ranking (BPR) \cite{rendle2009bpr}:
\begin{equation}
  \mathcal{L}_{\mathrm{BPR}}(s)
  = - \sum_{(u,i,j)} \log \sigma(s(u,i)-s(u,j)),
\end{equation}
where $i$ is an observed positive item and $j$ is a sampled negative
item. Existing multimodal recommenders usually define $s(u,i)$ by
making collaborative, visual, and textual embeddings consistent or
fusing them. In
this work, we make the scoring problem more explicit: content-derived
signals are allowed to affect ranking only when supported by behavior
evidence.

BRIDGE learns a final score of the form
\begin{equation}
  s(u,i)=s_{\mathrm{base}}(u,i)+\Delta_{\mathrm{bridge}}(u,i),
\end{equation}
where the residual correction is behavior-guided and applied only in a
controlled candidate scope. The concrete definitions of behavior
evidence and the frequency-aware encoder are given in Method.

\begin{table*}[!t]
  \caption{Overall recommendation performance on Amazon Baby, Sports,
  and Electronics. All methods use the same processed splits, BEIT3
  features, and full-sort evaluator. Entries report mean $\pm$ standard
  deviation over five seeds. $\ast\ast$ indicates $p < 0.01$ against the
  strongest baseline under an independent two-sample $t$-test.}
  \Description{A performance table compares BPR, graph-based, and
  multimodal recommendation baselines with BRIDGE on Amazon Baby,
  Sports, and Electronics using Recall and NDCG at 10 and 20. BRIDGE
  has the highest reported score for every dataset and metric.}
  \label{tab:main}
  \centering
  \scriptsize
  \setlength{\tabcolsep}{2pt}
  \resizebox{\textwidth}{!}{%
  \begin{tabular}{ll*{16}{c}}
  \toprule
  Dataset & Metric & BPR & LightGCN & VBPR & MMGCN & GRCN & DualGNN
  & SLMRec & LATTICE & BM3 & FREEDOM & DiffMM & MMIL & AlignRec
  & SMORE & FITMM & BRIDGE \\
  \midrule
  Baby & R@10 & 0.0357 & 0.0479 & 0.0418 & 0.0413 & 0.0538 & 0.0507
  & 0.0533 & 0.0561 & 0.0573 & 0.0624 & 0.0617 & 0.0670 & 0.0674$_{\pm0.0007}$
  & 0.0680$_{\pm0.0008}$ & 0.0716$_{\pm0.0006}$ & \textbf{0.0752}$_{\pm0.0004}^{\ast\ast}$ \\
  & R@20 & 0.0575 & 0.0754 & 0.0667 & 0.0649 & 0.0832 & 0.0782
  & 0.0788 & 0.0867 & 0.0904 & 0.0985 & 0.0978 & 0.1035 & 0.1046$_{\pm0.0010}$
  & 0.1035$_{\pm0.0011}$ & 0.1089$_{\pm0.0009}$ & \textbf{0.1128}$_{\pm0.0006}^{\ast\ast}$ \\
  & N@10 & 0.0192 & 0.0257 & 0.0223 & 0.0211 & 0.0285 & 0.0264
  & 0.0291 & 0.0305 & 0.0311 & 0.0324 & 0.0321 & 0.0361 & 0.0363$_{\pm0.0004}$
  & 0.0365$_{\pm0.0005}$ & 0.0387$_{\pm0.0005}$ & \textbf{0.0428}$_{\pm0.0003}^{\ast\ast}$ \\
  & N@20 & 0.0249 & 0.0328 & 0.0287 & 0.0275 & 0.0364 & 0.0335
  & 0.0365 & 0.0383 & 0.0395 & 0.0416 & 0.0408 & 0.0455 & 0.0458$_{\pm0.0005}$
  & 0.0457$_{\pm0.0006}$ & 0.0484$_{\pm0.0005}$ & \textbf{0.0525}$_{\pm0.0003}^{\ast\ast}$ \\
  \midrule
    Sports & R@10 & 0.0432 & 0.0569 & 0.0561 & 0.0394 & 0.0607 & 0.0574
    & 0.0667 & 0.0628 & 0.0659 & 0.0705 & 0.0687 & 0.0747 & 0.0758$_{\pm0.0008}$
    & 0.0762$_{\pm0.0009}$ & 0.0809$_{\pm0.0007}$ & \textbf{0.0860}$_{\pm0.0004}^{\ast\ast}$ \\
    & R@20 & 0.0653 & 0.0864 & 0.0857 & 0.0625 & 0.0928 & 0.0881
    & 0.0998 & 0.0961 & 0.0987 & 0.1077 & 0.1035 & 0.1133 & 0.1160$_{\pm0.0012}$
    & 0.1142$_{\pm0.0013}$ & 0.1187$_{\pm0.0010}$ & \textbf{0.1262}$_{\pm0.0007}^{\ast\ast}$ \\
    & N@10 & 0.0241 & 0.0311 & 0.0305 & 0.0203 & 0.0335 & 0.0316
    & 0.0366 & 0.0339 & 0.0357 & 0.0382 & 0.0357 & 0.0405 & 0.0414$_{\pm0.0005}$
    & 0.0408$_{\pm0.0005}$ & 0.0441$_{\pm0.0004}$ & \textbf{0.0490}$_{\pm0.0002}^{\ast\ast}$ \\
    & N@20 & 0.0298 & 0.0387 & 0.0386 & 0.0266 & 0.0421 & 0.0393
    & 0.0454 & 0.0431 & 0.0443 & 0.0478 & 0.0458 & 0.0505 & 0.0517$_{\pm0.0006}$
    & 0.0506$_{\pm0.0006}$ & 0.0538$_{\pm0.0005}$ & \textbf{0.0594}$_{\pm0.0004}^{\ast\ast}$ \\
  \midrule
  Electronics & R@10 & 0.0235 & 0.0363 & 0.0293 & 0.0207 & 0.0349 & 0.0363
  & 0.0392 & 0.0397 & 0.0437 & 0.0382 & 0.0418 & 0.0456 & 0.0472$_{\pm0.0008}$
  & 0.0474$_{\pm0.0009}$ & 0.0503$_{\pm0.0006}$ & \textbf{0.0549}$_{\pm0.0002}^{\ast\ast}$ \\
  & R@20 & 0.0367 & 0.0540 & 0.0458 & 0.0331 & 0.0529 & 0.0541
  & 0.0581 & 0.0573 & 0.0648 & 0.0588 & 0.0617 & 0.0673 & 0.0700$_{\pm0.0010}$
  & 0.0691$_{\pm0.0012}$ & 0.0725$_{\pm0.0009}$ & \textbf{0.0778}$_{\pm0.0004}^{\ast\ast}$ \\
  & N@10 & 0.0127 & 0.0204 & 0.0159 & 0.0109 & 0.0195 & 0.0202
  & 0.0223 & 0.0224 & 0.0247 & 0.0209 & 0.0224 & 0.0253 & 0.0262$_{\pm0.0004}$
  & 0.0261$_{\pm0.0004}$ & 0.0278$_{\pm0.0003}$ & \textbf{0.0326}$_{\pm0.0002}^{\ast\ast}$ \\
  & N@20 & 0.0161 & 0.0250 & 0.0202 & 0.0141 & 0.0241 & 0.0248
  & 0.0273 & 0.0271 & 0.0302 & 0.0262 & 0.0278 & 0.0309 & 0.0321$_{\pm0.0005}$
  & 0.0318$_{\pm0.0005}$ & 0.0335$_{\pm0.0004}$ & \textbf{0.0385}$_{\pm0.0003}^{\ast\ast}$ \\
  \bottomrule
  \end{tabular}}
\end{table*}

\section{Motivating Observations}

BRIDGE is motivated by three empirical observations. Together they
suggest a simple design principle: multimodal information should be
preserved, but content should be treated as conditional evidence, not
as a command.

\subsection{Direct Consistency Has an Effective Range}

We run a compact direct-consistency diagnostic on Amazon Baby using the
same backbone and split, while varying only the consistency strength.
Without direct consistency, Recall@20 is 0.0922. A moderate setting
with a contrastive loss weight of 0.01, a similarity weight of 0.3, and
UI cosine improves it to 0.1024, showing that cross-view consistency
can be useful. However, a stronger setting with a contrastive loss
weight of 0.1 and a similarity weight of 1.0 drops Recall@20 to 0.0836,
showing that direct consistency has an effective range. Once this
objective dominates, it can suppress private recommendation factors that
are not shared by all modalities. We also checked a lighter setting
with a contrastive loss weight of 0.001 and a similarity weight of 0.1
under another seed, which reached Recall@20 of 0.0982 and still stayed
below the moderate setting.

\subsection{Low Frequency Agrees; High Frequency Ranks}

\begin{figure*}[!t]
  \centering
  \includegraphics[width=0.88\textwidth]{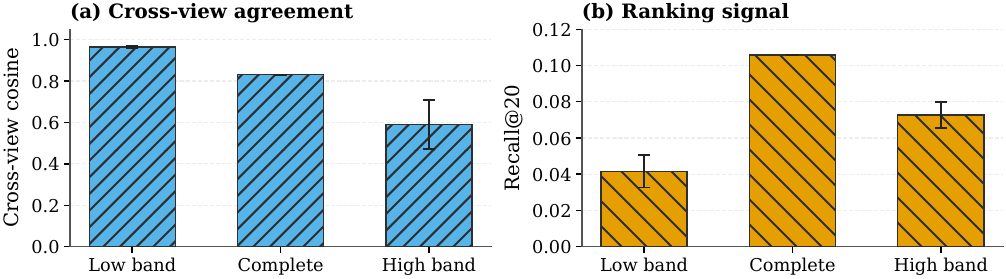}
  \caption{Representation-spectral frequency diagnostic on Amazon Baby.
  Low-frequency bands have stronger cross-view agreement, whereas
  high-frequency bands retain more ranking signal; the complete
  representation performs best. Error bars show the observed ranges.}
  \Description{Two bar charts compare low, complete, and high
  representation-spectral bands on Amazon Baby. Cross-view cosine is
  highest for the low band at 0.9645, lower for the complete
  representation at 0.8302, and lowest for the high band at 0.5894.
  Recall at 20 is 0.0416 for the low band, 0.1057 for the complete
  representation, and 0.0727 for the high band. Error bars show the
  observed ranges.}
  \label{fig:freqdiag-vis}
\end{figure*}

Frequency diagnostics show where the useful information sits.
Low-frequency components have stronger cross-view agreement, while
high-frequency components retain more ranking signal. Both are weaker
than the complete representation. Figure~\ref{fig:freqdiag-vis} shows
this comparison. We define these bands by ordering the singular
components of graph-smoothed channel matrices; Section~\ref{sec:dfge}
gives the formal construction. Low frequency captures shared structure,
whereas higher-frequency variation carries more local ranking evidence.

\subsection{Behavior Evidence Decides When Calibration Helps}

Content-derived corrections are not uniformly beneficial. In our
diagnostics, global score correction often hurts, while top-$K$ candidate
correction can improve ranking. Co-user behavior is most useful on the
sparser Amazon domains, so we use it as a local signal for candidate
calibration. A candidate supported by co-occurrence can receive a small
correction; otherwise, the same content relation may identify a
substitute or visual lookalike.

This yields three design requirements for BRIDGE:
\begin{enumerate}
  \item Preserve frequency-aware multimodal structure instead of
  collapsing all signals into an undifferentiated representation.
  \item Use behavior evidence inside the model so that user histories
  can authorize or reject candidate corrections.
  \item Restrict score calibration to the candidate set, so uncertain
  evidence cannot perturb the whole item space.
\end{enumerate}

\section{Method}

Figure~\ref{fig:overview} summarizes BRIDGE. The method preserves
frequency-specific multimodal evidence before using behavior to calibrate
the final ranking. DFGE learns the frequency-aware graph representation
and base score $s_{\mathrm{base}}(u,i)$. BEN estimates normalized
behavior evidence $b_{ui}$ from the training interaction graph. CRI
injects that evidence as a signed residual inside the candidate scope,
so the base encoder defines the ranking geometry and the behavior branch
adjusts local decisions.

\subsection{Dual-Frequency Graph Evidence Encoder}
\label{sec:dfge}

Each item has an ID embedding, a visual feature, and a textual feature.
The visual and textual features are projected to the same latent
dimension as the ID channel. DFGE first applies channel-specific
projection, then two rounds of symmetric normalized message passing on
the user-item graph, and finally sums the layer states to form each
channel embedding. Let \(\mathbf{A}_{ui}\) denote the user-item
adjacency and
\(\hat{\mathbf{A}}_{ui}=\mathbf{D}^{-\frac12}\mathbf{A}_{ui}\mathbf{D}^{-\frac12}\)
its symmetric normalization. For channel \(c\), the input state is
\(\mathbf{H}^{c,(0)}=[\mathbf{P}^{c}_u;\mathbf{W}_c\mathbf{x}^{c}_i]\),
where \(\mathbf{P}^{c}_u\) is a learned user preference table and
\(\mathbf{W}_c\) is a linear projection for that channel; for the ID
channel, the second term is the learned item ID embedding. We then
apply \(L=2\) rounds of normalized message passing,
\(\mathbf{H}^{c,(\ell+1)}=\hat{\mathbf{A}}_{ui}\mathbf{H}^{c,(\ell)}\),
and read out the channel embedding by summing all layers,
\(\mathbf{Z}^{c}=\sum_{\ell=0}^{L}\mathbf{H}^{c,(\ell)}\),
for \(c\in\{\mathrm{id},v,t\}\).
The item-side representation is refined with a cached multimodal item
graph built from top-$K$ cosine neighbors over the content features.
When both image and text are available, the cached item graph is the
weighted sum \(0.1\,\mathbf{A}_{v}+0.9\,\mathbf{A}_{t}\) with
\(K=10\). The item graph acts as a second smoothing stage before
frequency decomposition, using one propagation layer and keeping the
final smoothed item states rather than another layer-sum readout. This
produces user and item representations for the ID, visual, and textual
channels:
\begin{equation}
  \mathbf{z}^{c}_{u}, \mathbf{z}^{c}_{i}
  =
  \mathrm{GraphEnc}_{c}(\mathcal{R}, \mathbf{A}_{\mathrm{mm}},
  \mathbf{x}^{c}),
  \quad c \in \{\mathrm{id}, v, t\}.
\end{equation}
The three channels are then concatenated:
\begin{equation}
  \mathbf{z}_{u} =
  [\mathbf{z}^{\mathrm{id}}_{u};
   \mathbf{z}^{v}_{u};
   \mathbf{z}^{t}_{u}],
  \quad
  \mathbf{z}_{i} =
  [\mathbf{z}^{\mathrm{id}}_{i};
   \mathbf{z}^{v}_{i};
   \mathbf{z}^{t}_{i}].
\end{equation}
Each channel has dimension 64, so the fused representation dimension is
192.

\subsection{Spectral Band Routing}

DFGE decomposes each modality channel into $M$ frequency bands before
producing the base ranking embedding. For a representation matrix
$\mathbf{Z}^{c}$ of modality $c$, we compute a singular decomposition
$\mathbf{Z}^{c}=\mathbf{U}\Sigma\mathbf{V}^{\top}$ and split the
singular components into $M$ contiguous bands. The $m$-th band is:
\begin{equation}
  \mathbf{H}^{c,m}
  =
  \mathbf{U}_{m}\Sigma_{m}\mathbf{V}_{m}^{\top}.
\end{equation}
For each node $n$ and channel $c$, let
$\mathbf{h}^{c,m}_{n}=(\mathbf{H}^{c,m})_{n,:}$ denote the $m$-th band
slice. The band-specific multimodal state is then
\begin{equation}
  \mathbf{h}^{m}_{n}
  =
  [\mathbf{h}^{\mathrm{id},m}_{n};
   \mathbf{h}^{v,m}_{n};
   \mathbf{h}^{t,m}_{n}],
  \quad n \in \mathcal{U}\cup\mathcal{I}.
\end{equation}
A task-aware band gate then fuses the bands:
\begin{equation}
  \alpha^{m}_{n}
  =
  1 + \rho \cdot \phi_m(\mathbf{z}_{n}),
  \quad
  \omega_m = \sigma(a_m),
\end{equation}
\begin{equation}
  \mathbf{e}_{n}
  =
  \sum_{m=1}^{M} \omega_m \alpha^{m}_{n}\mathbf{h}^{m}_{n}.
\end{equation}
Here \(\phi(\mathbf{z}_n)=\sigma(\mathbf{W}_g\mathbf{z}_n+\mathbf{b}_g)\)
is a one-layer gate network with output dimension \(M\); in the main
setting, \(\phi_m(\cdot)\) is the \(m\)-th sigmoid output of this linear
gate, and \(\rho\) is a learned scalar initialized to \(0.5\).
The base score is computed by inner product:
\begin{equation}
  s_{\mathrm{base}}(u,i)
  =
  \langle \mathbf{e}_{u}, \mathbf{e}_{i} \rangle .
\end{equation}
The frequency decomposition is part of the base ranking representation.
Low-frequency bands, in the sense of leading singular directions of the
propagated representation spectrum, provide shared graph evidence,
whereas higher-frequency bands preserve private ranking variation. CRI
later decides whether a candidate score should receive a
behavior-authorized residual. The high-frequency bands remain part of the
base representation rather than a separate score for every item.

\paragraph{Interpretation of frequency.}
Here, frequency means the ordering of representation components after
graph smoothing. Let \(\mathbf{Z}\in\mathbb{R}^{N\times d}\) be a
graph-smoothed channel matrix after message passing. BRIDGE computes
\(\mathbf{Z}=\mathbf{U}\Sigma\mathbf{V}^{\top}\), partitions the right
singular directions as \(\mathbf{V}=[\mathbf{V}_1,\ldots,\mathbf{V}_M]\)
in descending singular-value order, and reconstructs band evidence by
\(\mathbf{H}^{m}=\mathbf{Z}\mathbf{V}_m\mathbf{V}_m^{\top}\). The leading
bands capture dominant post-smoothing variation, while later bands
collect residual directions with less shared energy. Because normalized
graph propagation is already low-pass, the leading singular bands act as
stable shared evidence and the trailing bands retain more local
variation. A graph Laplacian basis would order components by graph
oscillation, whereas this decomposition orders latent energy.

\paragraph{Channel-wise decomposition.}
The decomposition is applied to each modality channel before fusion.
Each user/item representation is split into
64-dimensional ID, image, and text blocks; each block is decomposed
separately and then reassembled into 192-dimensional band
representations. With $M=3$, the 64 singular directions are partitioned
into bands of size $22/21/21$ in descending singular-value order. For
controlled analysis, we also evaluate non-SVD decomposition operators
under the same encoder and fusion capacity. In the equal-capacity
no-decomposition control, the band inputs are
\begin{equation}
  \mathbf{H}^{c,m}=\frac{1}{M}\mathbf{Z}^{c},
  \quad m=1,\ldots,M.
\end{equation}
This keeps the same number of band gates, band weights, graph encoders,
and fusion parameters while removing spectral separation itself. We also
evaluate fixed-basis variants in which
$\mathbf{H}^{c,m}=\mathbf{Z}^{c}\mathbf{Q}_{m}\mathbf{Q}_{m}^{\top}$:
Gram uses eigenvectors of $\mathbf{Z}^{\top}\mathbf{Z}$, DCT uses a
deterministic cosine basis, and Random uses a fixed
orthogonal basis.

\subsection{Behavior Evidence Normalizer}

The Behavior Evidence Normalizer (BEN) builds an item-item behavior
graph from co-user similarity after the interaction split is fixed. Let
$\mathcal{D}_{\mathrm{tr}}$ denote
the training interactions only, and let $\mathcal{U}^{\mathrm{tr}}_i$
denote the users who interacted with item $i$ in
$\mathcal{D}_{\mathrm{tr}}$. The raw behavior edge between two items is:
\begin{equation}
  B_{ij} =
  \frac{|\mathcal{U}^{\mathrm{tr}}_i \cap
  \mathcal{U}^{\mathrm{tr}}_j|}
  {\sqrt{|\mathcal{U}^{\mathrm{tr}}_i|}
  \sqrt{|\mathcal{U}^{\mathrm{tr}}_j|}} .
\end{equation}
Only the top $K_b$ neighbors of each item are kept. For a user-item pair
$(u,i)$, the raw candidate behavior evidence is aggregated from the
user's training history $\mathcal{H}^{\mathrm{tr}}_{u}$:
\begin{equation}
  \widetilde{b}_{ui} =
  \frac{1}{\sqrt{|\mathcal{H}^{\mathrm{tr}}_u|}}
  \sum_{j \in \mathcal{H}^{\mathrm{tr}}_u} B_{ij}.
\end{equation}
Validation and test interactions are never used to construct
$B_{ij}$, $\mathcal{H}^{\mathrm{tr}}_{u}$, item popularity, or user
degree features; they are used only for model selection and final
evaluation.
The reported setting uses cosine co-user similarity, $K_b=1000$,
square-root history normalization, and centered behavior scores:
\begin{equation}
  b_{ui} =
  \frac{\widetilde{b}_{ui}-\mu_u}{\sigma_u+\epsilon}.
\end{equation}
Although $B_{ij}$ and $\widetilde{b}_{ui}$ are non-negative, the
normalized residual $b_{ui}$ can be positive or negative. Positive
values indicate stronger-than-usual behavior support under the user's
history, while negative values indicate weaker-than-usual support and
can demote a candidate when used in Eq.~\ref{eq:inference-score}.
When the full user-item behavior matrix fits in memory, BRIDGE
normalizes each user's complete behavior row once and reuses the same
$b_{ui}$ values in sampled training pairs and candidate inference. For
larger sparse datasets, $b_{ui}$ is computed on demand; both paths yield
the same normalized score and avoid hidden train--test mismatch.

\subsection{Candidate Residual Integrator}

The Candidate Residual Integrator (CRI) uses BEN evidence as a signed
score residual after DFGE has selected a candidate set. The central
design choice is the candidate-level scope rather than an unconstrained
learned coefficient. The main BRIDGE setting uses a validation-selected
residual strength:
\begin{equation}
  \Delta_{\mathrm{bridge}}(u,i)
  =
  \mathbb{I}[i \in \mathcal{C}^{\mathrm{tr}}_u]
  \lambda_b b_{ui}.
\end{equation}
This is the formulation used in the main experiments. We also include a
conservative control with lightweight user and item biases
$\beta_u,\beta_i$ and scalar coefficients $a_s,a_b$:
\begin{equation}
  \eta_{ui} =
  \sigma(
  \beta_u + \beta_i
  + a_s \tanh(s_{\mathrm{base}}(u,i))
  + a_b \tanh(b_{ui})
  ).
\end{equation}
The corresponding candidate score is
\begin{equation}
  s_{\mathrm{train}}(u,i) =
  s_{\mathrm{base}}(u,i)
  + \mathbb{I}[i \in \mathcal{C}^{\mathrm{tr}}_u]
  \lambda_b \eta_{ui}b_{ui}.
\end{equation}
In both cases, the sign of the correction comes from $b_{ui}$; the
coefficient only scales the residual. This distinction matters
empirically: the fixed-$\lambda_b$ calibrator is the primary CRI
instantiation, while the conservative coefficient remains a useful
control.
During training, BRIDGE uses the same candidate scope as inference. For
each training user, we first form a detached base-score candidate set:
\begin{equation}
  \mathcal{C}^{\mathrm{tr}}_{u} =
  \mathrm{TopK}(\mathrm{sg}(s_{\mathrm{base}}(u,\cdot)), K_c),
\end{equation}
where $\mathrm{sg}(\cdot)$ denotes stop-gradient. This set is computed
with the same candidate size $K_c$ used at inference. It is used only
as a scope mask, not as an additional supervision target. In the
reported setting, the set is recomputed for each mini-batch from
detached base embeddings rather than cached across epochs, so the
candidate scope tracks the evolving base scorer. Sampled
positive and negative items are then scored as:
\begin{equation}
  s_{\mathrm{train}}(u,i) =
  s_{\mathrm{base}}(u,i)
  + \mathbb{I}[i \in \mathcal{C}^{\mathrm{tr}}_u]
  \lambda_b b_{ui}.
  \label{eq:train-score}
\end{equation}
The indicator ensures that the residual branch is active only on
candidate items. Items outside $\mathcal{C}^{\mathrm{tr}}_u$ still
contribute through $s_{\mathrm{base}}(u,i)$, so the base scorer can
learn to move future positives into the candidate set. Gradients are
taken through the base score for all sampled pairs and through the
calibration coefficient only when the conservative control is used; the
top-$K$ membership itself is detached. This candidate-aware objective
reduces the train--test scope mismatch of training a residual on
globally sampled pairs while applying it only to a top-$K$ candidate
set at inference.
During inference, BRIDGE first obtains a candidate set from the base
multimodal score:
\begin{equation}
  \mathcal{C}_{u} =
  \mathrm{TopK}(s_{\mathrm{base}}(u,\cdot), K_c).
\end{equation}
At inference, the final score is:
\begin{equation}
  s(u,i) =
  s_{\mathrm{base}}(u,i)
  + \mathbb{I}[i \in \mathcal{C}_u]
  \lambda_b b_{ui}.
  \label{eq:inference-score}
\end{equation}
This top-$K$ restriction is the main operational scope. It lets behavior
evidence reorder plausible candidates, but prevents it from rewriting
the whole item space. The reported setting uses $K_c=200$, and
$\lambda_b$ is selected on validation Recall@20 from a small
calibration grid. The conservative coefficient shares the same score
form and candidate scope, but it is used only in the control variant.
BRIDGE therefore always performs behavior-residual candidate
calibration, and the calibration strength is selected on validation
rather than tuned on test.

\paragraph{Candidate-aware training.}
A simpler implementation
would train Eq.~\ref{eq:train-score} without the indicator and then
apply Eq.~\ref{eq:inference-score} only inside $\mathcal{C}_u$ at test
time. We treat that variant as a reranking baseline and compare against
it in the control suite. BRIDGE instead trains the residual under the same
scope used at inference, which keeps the residual branch focused on
candidate items and leaves the base scorer responsible for moving new
positives into the candidate set.

\subsection{Training Objective}

BRIDGE optimizes the calibrated BPR loss:
\begin{equation}
  \mathcal{L}_{\mathrm{rank}}
  =
  -\sum_{(u,i,j)}
  \log \sigma(s_{\mathrm{train}}(u,i)-s_{\mathrm{train}}(u,j)).
\end{equation}
An auxiliary BPR loss on the base score keeps the backbone predictive:
\begin{equation}
  \mathcal{L}_{\mathrm{base}}
  =
  -\sum_{(u,i,j)}
  \log \sigma(s_{\mathrm{base}}(u,i)-s_{\mathrm{base}}(u,j)).
\end{equation}
This auxiliary loss matters under candidate-aware training: when a
positive item is not yet in $\mathcal{C}^{\mathrm{tr}}_u$, the residual
branch is inactive, and the base loss supplies the learning signal that
can move the item into later candidate sets.
The frequency gates are regularized by two objectives: an
information-bottleneck surrogate that controls gate expansion, and a
frequency structural regularizer that constrains the decomposed bands.
Let $\gamma_{n,m}$ denote the frequency gate value for node $n$ and
band $m$, and define
$\delta_{n,m}=\max(\gamma_{n,m}-1,0)$ under the positive-gate setting.
The information-bottleneck surrogate is:
\begin{equation}
  \mathcal{L}_{\mathrm{IB}}
  =
  \frac{1}{|\mathcal{N}|}\sum_{n\in\mathcal{N}}
  \left[
  \alpha_{\mathrm{IB}}\|\boldsymbol{\delta}_{n}\|_{2}^{2}
  +
  \mu_{\mathrm{IB}}\|\boldsymbol{\delta}_{n}\|_{2}
  \sum_{m=1}^{M}\max(\delta_{n,m}-\phi_{+},0)
  \right],
  \label{eq:ib-loss}
\end{equation}
where $\mathcal{N}$ includes the user and item nodes used by the
encoder. This term penalizes unnecessary expansion of the frequency
gates: if a node can be represented without strongly amplifying a
particular frequency band, the gate is encouraged to stay close to one.

For a training mini-batch
$\mathcal{B}=\{(u_b,i_b,j_b)\}_{b=1}^{B}$, BRIDGE computes the
frequency structural regularizer from the user bands
$\mathbf{h}_{u_b}^{m}$ and the positive-item bands
$\mathbf{h}_{i_b}^{m}$. Let
$\tilde{\mathbf{h}}=\mathbf{h}/(\|\mathbf{h}\|_2+\epsilon)$ denote
the normalized band vector. The regularizer has two parts:
\begin{equation}
  \mathcal{L}_{\mathrm{freq}}
  =
  \mathcal{L}_{\mathrm{dec}}
  +
  \mathcal{L}_{\mathrm{disc}},
  \label{eq:freq-loss}
\end{equation}
where
\begin{equation}
\begin{split}
  \mathcal{L}_{\mathrm{dec}}
  =
  \frac{1}{B M(M-1)}
  \sum_{b=1}^{B}\sum_{m\neq \ell}
  &\left[
  \cos^{2}(\tilde{\mathbf{h}}^{m}_{u_b},\tilde{\mathbf{h}}^{\ell}_{u_b})
  \right. \\
  &\left.
  +
  \cos^{2}(\tilde{\mathbf{h}}^{m}_{i_b},\tilde{\mathbf{h}}^{\ell}_{i_b})
  \right],
  \label{eq:freq-dec-loss}
\end{split}
\end{equation}
and
\begin{equation}
\begin{split}
  \mathcal{L}_{\mathrm{disc}}
  =
  -\frac{1}{B|\mathcal{M}_{h}|}
  \sum_{m\in\mathcal{M}_{h}}
  \sum_{b=1}^{B}
  \log
  \frac{
  \exp(\langle \tilde{\mathbf{h}}^{m}_{u_b},\tilde{\mathbf{h}}^{m}_{i_b}\rangle/\tau)}
  {\sum_{r=1}^{B}\exp(\langle \tilde{\mathbf{h}}^{m}_{u_b},\tilde{\mathbf{h}}^{m}_{i_r}\rangle/\tau)}.
  \label{eq:freq-disc-loss}
\end{split}
\end{equation}
Here $\mathcal{M}_{h}$ denotes the higher-frequency bands. The
decorrelation term suppresses redundant overlap between different bands,
while the high-frequency discrimination term keeps the higher-frequency
bands ranking-aware by contrasting each positive pair against other
positive items in the same batch. This regularizer is structural rather
than a strict orthogonality penalty: it preserves a stable band geometry
while keeping the informative bands discriminative. The IB surrogate
controls how much each node may amplify frequency bands, whereas the
frequency structural regularizer controls the geometry of the resulting
band embeddings.
When the conservative coefficient control is enabled, we add a soft
constraint on the average gate value:
\begin{equation}
  \mathcal{L}_{\eta}
  =
  \left(\mathbb{E}_{(u,i)}[\eta_{ui}] - \tau_{\eta}\right)^2.
\end{equation}
For the main fixed-$\lambda_b$ BRIDGE setting, $\mathcal{L}_{\eta}=0$.
The final objective is:
\begin{equation}
\begin{split}
  \mathcal{L} =
  &\mathcal{L}_{\mathrm{rank}}
  + \lambda_{\mathrm{base}}\mathcal{L}_{\mathrm{base}}
  + \lambda_{\mathrm{IB}}\mathcal{L}_{\mathrm{IB}}
  + \lambda_{\mathrm{freq}}\mathcal{L}_{\mathrm{freq}} \\
  &+ \lambda_{\eta}\mathcal{L}_{\eta}.
\end{split}
\end{equation}
In the reported fixed-$\lambda_b$ setting, the active coefficients are
\(\lambda_{\mathrm{base}}=0.2\),
\(\lambda_{\mathrm{IB}}=1.0\),
\(\lambda_{\mathrm{freq}}=0.001\), and
\(\lambda_{\eta}=0\). The conservative coefficient control uses
\(\lambda_{\eta}=0.001\) and gate target \(\tau_{\eta}=0.5\). The
candidate-calibration strength follows the same five-point validation
grid.
This objective makes the method explicit: DFGE learns the
frequency-organized multimodal representation, and CRI integrates BEN
evidence under the same candidate scope used by inference.

\section{Experiments}

\subsection{Experimental Setup}

We evaluate BRIDGE on Amazon Baby, Sports, and Electronics from the
Amazon review benchmark \cite{mcauley2015image} under the common
MMRec-style implicit-feedback split. Baby has 19,445 users, 7,050
items, and 160,792 interactions. Sports has 35,598 users, 18,357
items, and 296,337 interactions. Electronics has 192,403 users,
63,001 items, and 1,689,188 interactions. We use BEIT3-derived
multimodal features and report Recall@10, Recall@20, NDCG@10, and
NDCG@20. All baselines in Table~\ref{tab:main}
use the same processed interaction splits, BEIT3 feature files, and
full-sort evaluation protocol. Unless otherwise stated, reported
numbers use seed 2020.

For all three datasets, the behavior graph is constructed strictly from
the training split. Each interaction file carries a
training, validation, or test partition marker. We first build
$\mathcal{D}_{\mathrm{tr}}$ from the training partition and compute
co-user edges, user histories, item popularity, and user degree from
this training matrix only. Validation and test interactions are not
used in the behavior graph. Baby contains 118,551 training
interactions, 20,559 validation interactions, and 21,682 test
interactions. Sports contains 218,409 training interactions, 37,899
validation interactions, and 40,029 test interactions. Electronics
contains 1,254,441 training interactions, 211,296 validation
interactions, and 223,451 test interactions. After filtering, validation
and test contain no cold-start users.

All BRIDGE main results use candidate-aware training. For each training
mini-batch, the model computes the detached base top-$K_c$ candidate set
of each involved user and masks the behavior residual outside this set,
as defined in Eq.~\ref{eq:train-score}. At test time, the same
$K_c$-restricted residual is applied to the base top-$K_c$ list in
Eq.~\ref{eq:inference-score}. Training uses pairwise BPR with this same
candidate mask, so the residual is learned in the scope where it is used.

The reported hyperparameters follow the dataset and model configuration
files. We use a 64-dimensional embedding size, two graph layers, and
three frequency bands. The item graph uses 10 neighbors. The learning
rate is $10^{-4}$, the regularization weight is $10^{-3}$, the IB
weight is 1.0, the behavior top-$K$ is 1000, the behavior weight is
0.4, the behavior evaluation top-$K$ is 200, and the train candidate
top-$K$ is 200. The behavior graph uses cosine similarity, sum-sqrt
aggregation, and z-score normalization. Training uses Adam for 300
epochs with batch sizes 2048 and 4096, zero weight decay, and seed
2020. These settings are shared by the main tables, the ablation
suite, and the calibration sweeps. The experiment suite is organized to
match the method: Table 1 compares overall ranking quality, Tables 2
and 3 isolate the residual and spectral controls, and Table 4 checks
whether the gains come from a head-item boost or from a broader
redistribution of ranking mass.

\subsection{Main Results}

Table~\ref{tab:main} follows the common overall comparison format and
places BRIDGE in the rightmost column. The baseline set spans pairwise
ranking (BPR \cite{rendle2009bpr}) and graph collaborative models
(LightGCN \cite{he2020lightgcn}, MMGCN \cite{wei2019mmgcn}, GRCN
\cite{wei2020grcn}, DualGNN \cite{wang2021dualgnn}, and LATTICE
\cite{zhang2021lattice}). It also includes multimodal variants (VBPR
\cite{he2016vbpr}, SLMRec \cite{tao2022slmrec}, BM3
\cite{zhou2023bm3}, FREEDOM \cite{zhou2023freedom}, DiffMM
\cite{jiang2024diffmm}, MMIL \cite{yang2024mmil}, AlignRec
\cite{liu2024alignrec}, SMORE \cite{ong2024smore}, and FITMM
\cite{yang2025fitmm}). The comparison therefore covers classical
collaborative ranking and recent multimodal systems under the same
data pipeline.
We select the checkpoint and candidate-calibration setting by
validation Recall@20. The validation grid sets $\lambda_b$ to 0.1,
0.2, 0.4, 0.6, or 0.8, and also evaluates a coefficient-controlled
variant. The test set is evaluated only after this selection. On the
three reported datasets, the selected BRIDGE setting is the
fixed-$\lambda_b$ candidate calibration rather than the conservative
coefficient control.
For the recent strong baselines and BRIDGE, we report mean and standard
deviation across five seeds because these methods define the competitive
margin. The remaining baselines are included as single-run protocol
controls under the same processed data and feature setting.
Hyperparameters are chosen on validation Recall@20 using the ranges
provided by the corresponding implementations or MMRec-style
configuration files. Table~\ref{tab:main} shows that BRIDGE improves
every metric on every dataset. The Recall gains are consistent, but the
NDCG gains are larger, which means the method mainly sharpens the order
near the top of the list rather than only recovering more positives at
larger cutoffs. The margin is most pronounced on Electronics, where
sparsity and popularity skew make a global residual less reliable. This
matches the design of BRIDGE: DFGE sets the candidate geometry, and
BEN/CRI refines only the items that already look plausible under the
multimodal backbone.
Broken down by dataset, Baby shows the smallest but still stable gain,
Sports shows the largest balanced improvement across Recall and NDCG,
and Electronics shows the clearest benefit from candidate-level
calibration. This pattern is consistent with the data regime: as the
interaction graph becomes sparser, the model benefits more from a
restricted residual than from a global score adjustment.

\subsection{Diagnostics}

At the validation-selected Baby checkpoint, the pair coefficient mean is
0.7365, the behavior correction absolute mean is 0.6886, the low-item
norm mean is 6.3819, the high-item norm mean is 2.7333, and the
low-high item cosine is 0.0560. These values show that the two frequency
components capture different signals and that the residual branch acts
locally. The low-frequency band carries stronger cross-view agreement
and larger energy, while the high-frequency band is less aligned but
remains informative. Their near-zero cosine indicates that BRIDGE can
use the low bands as shared structure and the high bands as local
variation.

\subsection{Additional Controls and Sensitivity}
\label{sec:controls}

Tables~\ref{tab:ablation}--\ref{tab:popularity-exposure} report the
controls that separate the roles of DFGE, BEN, and CRI.
All controls use the same splits, features, and full-sort protocol.
The control suite answers three questions. First, do BEN and CRI need
behavior evidence and candidate restriction, or can a generic residual
work? Second, does DFGE need a spectral decomposition, or is
equal-capacity routing enough? Third, does the residual mainly move
score mass away from popular items, or does it improve ranking across
strata? Tables~\ref{tab:ablation}--\ref{tab:popularity-exposure}
provide the quantitative controls.

\begin{table*}[!t]
  \caption{Core Baby ablations and cross-dataset residual alternatives.
  All rows use the same splits, BEIT3 features, candidate restriction,
  and full-sort evaluation protocol.}
  \Description{A two-part ablation table on Amazon Baby shows that
  removing behavior evidence or top-K calibration lowers BRIDGE's
  performance, while frequency variants produce smaller changes. A
  second part compares raw co-occurrence, EASE, conservative
  coefficients, and BRIDGE across three datasets; BRIDGE is best in the
  listed Recall at 20 and NDCG at 20 results.}
  \label{tab:ablation}
  \centering
  \scriptsize
  \setlength{\tabcolsep}{3pt}
  \begin{tabular*}{\textwidth}{@{\extracolsep{\fill}}llcccc}
    \toprule
    \multicolumn{6}{l}{\textit{(a) DFGE, BEN/CRI, and frequency ablations on Amazon Baby}} \\
    \midrule
    Module & Variant & R@10 & R@20 & N@10 & N@20 \\
    \midrule
    Full & BRIDGE & \textbf{0.0752} & \textbf{0.1128} & \textbf{0.0428} & \textbf{0.0525} \\
    \midrule
    BEN/CRI & w/o behavior evidence & 0.0682 & 0.1016 & 0.0368 & 0.0453 \\
    BEN/CRI & w/o top-$K$ calibration & 0.0651 & 0.1025 & 0.0354 & 0.0450 \\
    BEN/CRI & global correction & 0.0745 & 0.1092 & 0.0419 & 0.0507 \\
    \midrule
    DFGE-Content & w/o item graph & 0.0535 & 0.0824 & 0.0282 & 0.0356 \\
    DFGE-Content & w/o image feature & 0.0735 & 0.1089 & 0.0415 & 0.0506 \\
    DFGE-Content & w/o text feature & 0.0750 & 0.1089 & 0.0422 & 0.0509 \\
    DFGE-Content & w/o content features & 0.0608 & 0.0894 & 0.0340 & 0.0414 \\
    \midrule
    DFGE-Spectral & w/o IB surrogate & 0.0709 & 0.1031 & 0.0404 & 0.0487 \\
    DFGE-Spectral & w/o freq. structural reg. & 0.0747 & 0.1107 & 0.0424 & 0.0517 \\
    DFGE-Spectral & w/o both freq. objectives & 0.0712 & 0.1023 & 0.0407 & 0.0487 \\
    DFGE-Spectral & w/o freq. decomp. (equal cap.) & 0.0746 & 0.1110 & 0.0425 & 0.0519 \\
    DFGE-Spectral & Gram decomposition & 0.0752 & 0.1117 & 0.0425 & 0.0519 \\
    DFGE-Spectral & DCT decomposition & 0.0751 & 0.1089 & 0.0427 & 0.0514 \\
    DFGE-Spectral & Random orthogonal decomp. & 0.0750 & 0.1113 & 0.0427 & 0.0519 \\
    \midrule
    \multicolumn{6}{l}{\textit{(b) Candidate-level alternative residual controls}} \\
    \midrule
    Dataset & Metric & Raw cooc. & EASE & Conserv. coeff. & BRIDGE \\
    \midrule
    Baby & R@20 & $0.1067$ & $0.1087$ & $0.1126$ & $\mathbf{0.1128}$ \\
    & N@20 & $0.0478$ & $0.0494$ & $0.0519$ & $\mathbf{0.0525}$ \\
    Sports & R@20 & $0.1218$ & $0.1229$ & $0.1240$ & $\mathbf{0.1262}$ \\
    & N@20 & $0.0553$ & $0.0571$ & $0.0589$ & $\mathbf{0.0594}$ \\
    Electronics & R@20 & $0.0713$ & $0.0724$ & $0.0738$ & $\mathbf{0.0778}$ \\
    & N@20 & $0.0338$ & $0.0344$ & $0.0351$ & $\mathbf{0.0385}$ \\
    \bottomrule
  \end{tabular*}
\end{table*}

\begin{table*}[!t]
  \caption{Candidate-side controls on Amazon Baby, Sports, and
  Electronics. To keep the control study in the main text, we report
  the primary Recall@20 and NDCG@20 metrics.}
  \Description{Two tables compare cross-dataset ablations and learned
  fusion controls. Restricting corrections to the top-K candidate set
  and retaining behavior evidence improves results over global or
  removed-calibration controls on Sports and Electronics.}
  \label{tab:candidate-controls}
  \centering
  \scriptsize
  \setlength{\tabcolsep}{2.5pt}
  \begin{minipage}[t]{0.51\textwidth}
    \centering
    \textit{(a) Cross-dataset ablations} \\
    \begin{tabular*}{\linewidth}{@{\extracolsep{\fill}}lp{31mm}cc}
      \toprule
      Dataset & Variant & R@20 & N@20 \\
      \midrule
      Sports & \textbf{BRIDGE} & $\mathbf{0.1262}$ & $\mathbf{0.0594}$ \\
      & conserv. coeff. & $0.1217$ & $0.0572$ \\
      & w/o behavior evidence & $0.1162$ & $0.0515$ \\
      & w/o top-$K$ calibration & $0.1127$ & $0.0493$ \\
      & global correction & $0.1137$ & $0.0542$ \\
      & w/o calib. coeff. & $0.1194$ & $0.0567$ \\
      & w/o behavior coeff. & $0.1236$ & $0.0581$ \\
      & w/o image feature & $0.1208$ & $0.0567$ \\
      & w/o text feature & $0.1213$ & $0.0572$ \\
      & w/o content features & $0.1048$ & $0.0505$ \\
      & w/o item graph & $0.1010$ & $0.0471$ \\
      \midrule
      Electronics & \textbf{BRIDGE} & $\mathbf{0.0778}$ & $\mathbf{0.0385}$ \\
      & conserv. coeff. & $0.0747$ & $0.0354$ \\
      & w/o calib. coeff. & $0.0755$ & $0.0359$ \\
      & w/o behavior evidence & $0.0656$ & $0.0298$ \\
      & w/o top-$K$ calibration & $0.0647$ & $0.0292$ \\
      & global correction & $0.0642$ & $0.0303$ \\
      & w/o image feature & $0.0648$ & $0.0293$ \\
      \bottomrule
    \end{tabular*}
  \end{minipage}\hfill
  \begin{minipage}[t]{0.46\textwidth}
    \centering
    \textit{(b) Learned fusion controls} \\
    \begin{tabular*}{\linewidth}{@{\extracolsep{\fill}}lp{30mm}cc}
      \toprule
      Dataset & Variant & R@20 & N@20 \\
      \midrule
      Baby & base encoder & $0.1017$ & $0.0455$ \\
      & fixed $\lambda$ mix & $0.1128$ & $0.0520$ \\
      & learned $\lambda_u$ & $0.1126$ & $0.0519$ \\
      & MLP gate & $0.1115$ & $0.0512$ \\
      & EASE residual & $0.1087$ & $0.0494$ \\
      & \textbf{BRIDGE} & $\mathbf{0.1128}$ & $\mathbf{0.0525}$ \\
      \midrule
      Sports & base encoder & $0.1141$ & $0.0502$ \\
      & \textbf{BRIDGE} & $\mathbf{0.1262}$ & $0.0594$ \\
      & learned $\lambda_u$ & $0.1260$ & $\mathbf{0.0595}$ \\
      & MLP gate & $0.1257$ & $0.0581$ \\
      & conserv. coeff. & $0.1240$ & $0.0589$ \\
      \bottomrule
    \end{tabular*}
  \end{minipage}
\end{table*}

\begin{table*}[!t]
  \caption{Popularity-stratified diagnostic and exposure track. Head,
  Mid, Tail, and Cold report Recall@20 within each item-popularity
  stratum; Head Exp. is the top-20 exposure share assigned to head
  items.}
  \Description{A popularity-stratified table compares base, control,
  and BRIDGE variants on Baby, Sports, and Electronics. BRIDGE improves
  total, mid, tail, and cold Recall at 20 while reducing the top-20
  exposure share assigned to head items relative to the base encoder.}
  \label{tab:popularity-exposure}
  \centering
  \scriptsize
  \setlength{\tabcolsep}{3pt}
  \begin{tabular*}{\textwidth}{@{\extracolsep{\fill}}llcccccc}
    \toprule
    Dataset & Variant & R@20 & Head & Mid & Tail & Cold & Head Exp. \\
    \midrule
    Baby & \quad base & $0.1017$ & $0.1649$ & $0.0094$ & $0.0059$ & $0.0031$ & $0.9842$ \\
    & \quad mix & $0.1110$ & $0.1705$ & $0.0270$ & $0.0183$ & $0.0109$ & $0.9356$ \\
    & \quad logic & $0.1113$ & $0.1710$ & $0.0270$ & $0.0189$ & $0.0109$ & $0.9358$ \\
    & \textbf{BRIDGE (fixed $\lambda$)} & $\mathbf{0.1128}$ & $\mathbf{0.1739}$ & $0.0261$ & $\mathbf{0.0189}$ & $\mathbf{0.0109}$ & $0.9426$ \\
    \midrule
    Sports & \quad base & $0.1141$ & $0.1888$ & $0.0338$ & $0.0217$ & $0.0119$ & $0.9073$ \\
    & \textbf{BRIDGE (fixed $\lambda$)} & $\mathbf{0.1262}$ & $\mathbf{0.1986}$ & $\mathbf{0.0497}$ & $\mathbf{0.0438}$ & $\mathbf{0.0283}$ & $0.8429$ \\
    & \quad conservative coeff. variant & $0.1240$ & $0.1926$ & $0.0452$ & $0.0381$ & $0.0231$ & $0.8658$ \\
    \midrule
    Electronics & \quad base & $0.0637$ & $0.0933$ & $0.0027$ & $0.0012$ & $0.0006$ & $0.9939$ \\
    & \textbf{BRIDGE (fixed $\lambda$)} & $\mathbf{0.0778}$ & $\mathbf{0.1115}$ & $\mathbf{0.0099}$ & $\mathbf{0.0045}$ & $\mathbf{0.0024}$ & $0.9726$ \\
    & \quad conservative coeff. variant & $0.0738$ & $0.1067$ & $0.0067$ & $0.0029$ & $0.0016$ & $0.9858$ \\
    \bottomrule
  \end{tabular*}
\end{table*}

Table~\ref{tab:ablation} isolates the behavior branch, multimodal
encoder, and spectral objectives on Baby. Panel (b) compares BRIDGE with
alternative residual rules on all three datasets, separating the score
itself from the way it is applied.

The first row block in Table~\ref{tab:ablation}(a) shows that BEN and
CRI are tied together. Removing behavior evidence hurts, but keeping
the same evidence and relaxing the scope is still worse than the full
candidate-aware model. Global correction recovers part of the drop but
remains below BRIDGE because the residual then reaches items outside the
base candidate set. The residual helps when it stays local.

The content block in Table~\ref{tab:ablation}(a) shows that the item
graph carries the largest share of the multimodal gain. Removing the
item graph causes a sharp drop, while removing image or text alone
produces smaller losses, which means the two modalities are
complementary rather than interchangeable. The spectral block then
separates architecture from regularization. Disabling the IB surrogate
or the frequency regularizer hurts performance, but Gram, DCT, and
random bases stay close to one another. The result points to a stable
band split as the important factor, rather than a particular basis.

Table~\ref{tab:ablation}(b) compares the residual against simpler
baselines.
Raw co-occurrence is the weakest control, EASE improves it, and the
conservative coefficient closes part of the gap. BRIDGE still wins on
all three datasets, with the largest margin on Electronics. The
comparison shows that BEN is more than an item-item prior. Its
normalization and candidate scope matter when the interaction matrix is
sparse and the residual must discriminate among near ties.

Table~\ref{tab:candidate-controls} extends the same check across
datasets and fusion rules. The cross-dataset block repeats the Baby
ordering on Sports and Electronics, so the mechanism is not tied to one
domain density. The learned-fusion block then
compares a fixed residual with learned scalar mixing, an MLP gate, and
an EASE-style residual. The learned variants remain close, but none
surpass the fixed BRIDGE setting. A larger fusion head adds little
beyond the candidate-restricted residual.

Table~\ref{tab:popularity-exposure} tests whether the gain comes from
favoring head items. BRIDGE raises Recall@20 in every
popularity stratum while reducing head exposure, and the effect is
strongest on Sports and Electronics, where the base model is more
skewed toward popular items. The conservative coefficient variant keeps
more of the original exposure pattern and usually trails the fixed
residual, making the fixed calibration rule the stronger operating point.

\begin{figure}[!htbp]
  \centering
  \includegraphics[width=\linewidth]{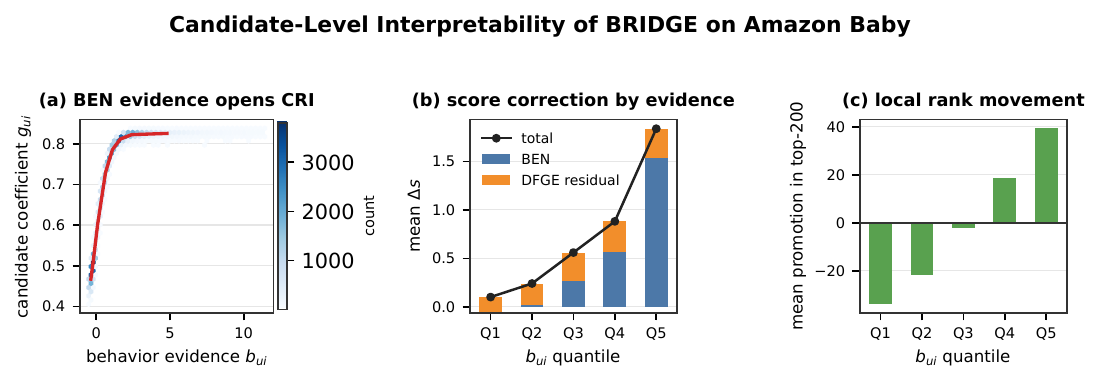}
  \includegraphics[width=\linewidth]{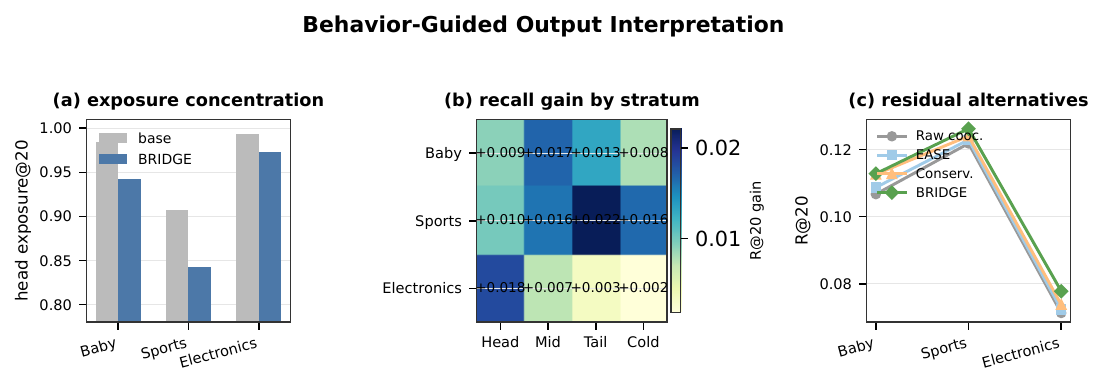}
  \vspace{2pt}
  \caption{Candidate-level interpretability controls for BRIDGE. The
  upper row shows BEN evidence, score corrections, and rank changes; the
  lower row shows exposure shifts, stratum-level recall, and residual
  alternatives.}
  \Description{The upper panel shows that the candidate coefficient rises
  quickly with behavior evidence, that positive evidence produces larger
  score corrections, and that high-evidence candidates gain rank within
  the top 200. The lower panel shows that BRIDGE reduces head-item
  exposure on Baby, Sports, and Electronics while improving recall in
  head, middle, tail, and cold strata. A final line chart compares raw
  co-occurrence, EASE, a conservative variant, and BRIDGE; BRIDGE is the
  strongest or tied strongest residual alternative across the three
  datasets.}
  \label{fig:bridge-controls}
\end{figure}

Across these controls, the same pattern appears. Removing behavior
evidence or candidate restriction hurts, and global correction is weaker
than candidate-level calibration. Raw co-occurrence and EASE do not
match BRIDGE's normalized residual, so BEN is more than an ItemKNN-style
lookup. The item graph and multimodal content shape candidate geometry,
while the spectral objectives stabilize the calibrated score.
Table~\ref{tab:popularity-exposure} shows the output effect: BRIDGE
raises mid, tail, and cold recall while lowering head exposure.

\section{Discussion and Limitations}

The controls separate the two roles in BRIDGE. Candidate restriction
keeps the residual from distorting the base score: removing the top-$K$
scope or applying the correction globally weakens ranking. Raw
co-occurrence, EASE, and learned fusion controls also remain below the
full model. The popularity results add an output-level view: the
residual improves all item strata while reducing head exposure, without
an explicit anti-popularity objective.

\subsection{Practical Implications}

DFGE builds the candidate manifold, BEN converts co-user overlap into
signed evidence, and CRI decides where that evidence may alter the
score. Keeping these roles separate is what makes the residual useful.
When the same evidence acts globally, it competes with the multimodal
backbone and the improvement shrinks. Candidate restriction keeps the
residual tied to the ranking decision.

The popularity diagnostics follow the same division of labor. Head items
are easier to reconstruct from sparse co-occurrence, while mid, tail,
and cold items depend more on cross-modal structure. BRIDGE uses no
explicit anti-popularity penalty. It preserves head accuracy and sends
more ranking mass to less exposed items that pass the candidate gate.
The exposure shift in Table~\ref{tab:popularity-exposure} follows from
this localization.

At inference, BRIDGE applies a residual pass to a cached candidate list.
The item graph is precomputed, the behavior graph uses training
interactions only, and the residual changes scores inside the shortlist.
This fits a standard retrieval-and-rank pipeline. Candidate quality sets
the upper bound: a relevant item that never enters the shortlist cannot
be recovered by calibration.

Noise in co-user neighborhoods blurs BEN when overlap is sparse or
interaction bursts are highly local. The conservative coefficient limits
the correction in these cases. The fixed candidate-weighted residual is
stronger once the backbone produces a reasonable shortlist, which is the
setting evaluated in the benchmark.

\subsection{Calibration Behavior}

The validation sweep gives the same qualitative behavior across the
coefficient settings. The fixed-$\lambda_b$ setting wins on all three
datasets, while the conservative control stays close. The residual is a
local correction, so a small coefficient has little effect and a large
one competes with the backbone. The five-point grid selects the middle
range without a large search.

The same pattern appears across the frequency design. The
low-frequency component carries the stable cross-view structure, while
the higher-frequency bands capture sparse and item-specific variation.
The IB surrogate and the frequency regularizer are both needed because
they serve different roles: one controls how much each node can route
into the bands, the other preserves a useful band geometry once the
routing has happened. The ablations show that dropping either term
weakens the ranking signal, while changing the basis has only a small
effect. The decomposition is structural and largely independent of the
basis.

Removing frequency decomposition gives 0.1110 Recall@20 on Baby versus
0.1128 for BRIDGE, while Gram, DCT, and random orthogonal bases remain
close. The gain therefore comes from separating shared low-frequency
evidence from more private higher-frequency variation; band routing and
its regularizers provide a smaller complementary gain by shaping the
candidate geometry used by BEN/CRI.

\subsection{Scope and Reproducibility}

BRIDGE operates at the ranking stage of a two-stage recommender. The
multimodal backbone builds the base top-$K$ list, which defines the
candidate universe, and BEN/CRI changes only the order inside that list.
Recall captures candidate coverage, whereas NDCG also reflects the
position of a relevant item in the local ranking. The larger NDCG gains in
Table~\ref{tab:main} arise from advancing plausible items within the list,
especially on the sparser Electronics graph.

Figure~\ref{fig:bridge-controls} makes the mechanism visible at the
candidate level. Positive BEN values produce larger score adjustments and
generally move the associated items upward in the top-200 order. The
dataset-level diagnostics follow the same pattern:
Table~\ref{tab:popularity-exposure} reports higher Recall@20 in the mid,
tail, and cold strata alongside lower head-item exposure on Sports and
Electronics. The residual reallocates positions among candidates that have
already passed the multimodal gate.

The fixed residual keeps this calibration step transparent. The base
representation supplies the retrieval signal, while normalized co-user
evidence resolves close scores. Learned scalar mixing and the MLP gate
remain competitive in Table~\ref{tab:candidate-controls}, yet the fixed
rule gives the strongest overall balance across the reported datasets. A
five-point validation grid sets its strength, keeping the final ranking
step easy to inspect and reproduce.

The ablations also map the design to deployment choices: candidate size
sets recall headroom, band count controls spectral resolution, and residual
strength controls how far behavior evidence can move a candidate. Baby
shows a modest calibration lift, Sports benefits from denser co-user
overlap, and Electronics gains the most NDCG from resolving sparse local
ties. The residual therefore complements the backbone rather than
replacing retrieval.

BRIDGE is most useful when the backbone supplies a semantically
plausible shortlist and local ordering is uncertain. Sparse overlap and
highly local interaction bursts make BEN less precise, and the residual
affects only items retained by the base candidate list. The study covers
three Amazon domains with one BEIT3 feature type. The item and behavior
graphs are cached and CRI evaluates only the shortlist; end-to-end latency
and throughput are not evaluated.

The public repository includes the model and evaluation code, dataset
format documentation, configurations, and run scripts for all three
benchmarks. Every experiment uses the same feature cache, split manifest,
and full-sort evaluator. Behavior evidence is constructed from the
training split, and the residual strength is selected on validation data.
These settings keep the ranking protocol aligned across all compared
methods.

The release also records candidate size, behavior top-$K$, band count,
random seed, feature paths, preprocessing scripts, and evaluation
commands. Keeping these choices in the configuration files makes it
possible to regenerate the tables with the same cached graphs and feature
files, while varying one component at a time.

\section{Conclusion}

BRIDGE separates candidate retrieval from local ranking. DFGE builds the
multimodal candidate geometry, BEN turns co-user overlap into signed
evidence, and CRI uses that evidence only to reorder candidates already
admitted by the backbone. Across Baby, Sports, and Electronics, this
division improves both Recall and NDCG, with the largest gains in NDCG.
Multimodal evidence finds plausible items; behavior evidence resolves their
close ordering.

The ablations support this division. BRIDGE changes local order without
changing the candidate pool, which explains the larger NDCG gains and lower
head-item exposure on Sports and Electronics.

The fixed residual works because broad preference matching stays with the
base score, while signed behavior evidence is reserved for close
alternatives. Its influence is concentrated near the top of the ranked list,
where small ordering changes matter most.

The tuning rule follows directly: use the backbone for candidate coverage and
apply the residual only after the shortlist is reliable. This separation
makes changes easier to diagnose: recall changes point to candidate formation,
whereas ranking changes point to local calibration.

This separation also explains the metric pattern: Recall changes only when
candidate coverage changes, whereas NDCG can improve when a relevant item is
moved upward within the same shortlist. The reported gains therefore measure
local ranking quality in addition to retrieval quality.

\begin{acks}
This work was supported by the
\grantsponsor{GS501100001809}{National Natural Science Foundation of
China}{https://doi.org/10.13039/501100001809} under Grant
No.~\grantnum{GS501100001809}{62271466}.
\end{acks}

\clearpage

\section*{GenAI Usage Disclosure}
The authors used generative AI tools for language editing, \LaTeX{}
formatting, drafting assistance, and figure preparation. The authors
developed and verified all technical content, experiments, analyses,
and conclusions and take responsibility for the final manuscript.

\bibliographystyle{ACM-Reference-Format}
\bibliography{bridge-cikm26-refs}

\end{document}